\documentclass[12pt,keywords,subjects]{notarticle}
\usepackage[dvips]{graphicx}

\begin{document}
\title{Partonometry in W + jet production}
\author{James Amundson\footnote{email: amundson@pa.msu.edu} 
    \and Jon Pumplin\footnote{email: pumplin@pa.msu.edu} 
    ~and Carl Schmidt\footnote{email: schmidt@pa.msu.edu}}
\address{Michigan State University \\ East Lansing, Michigan 48824}
\documentid{hep-ph/9708458}
\documentid{MSU-HEP-70822}
\date{\today}
\maketitle

\begin{abstract}
QCD predicts soft radiation patterns that are particularly simple for
$W+ \mbox{jet}$ production. We demonstrate how these patterns can be
used to distinguish between the parton-level subprocesses
probabilistically on an event-by-event basis. As a test of our method
we demonstrate correlations between the soft radiation and the
radiation inside the outgoing jet.
\end{abstract}

\section {Introduction}
\label{sec:intro}
Hard scattering in QCD is accompanied by soft radiation of  gluons
caused by the drastic accelerations of color charge.  The  resulting
antenna patterns are predicted to be particularly simple  for $2 \to
2$ hard scattering processes in which one of the final  particles is
colorless \cite{khoze,dokshitzer}.  The possibility  of using the
radiation pattern to identify the underlying  parton structure of
observed events has been called  ``partonometry'' \cite{khoze}.  In
this paper, we address the  question of how partonometry can be
carried out in practice,  and how it can be used to make further
studies of QCD.

As a specific example that is accessible to experiment, we study the
process $\bar p p \to W^\pm + {\rm jet}$ at the Fermilab Tevatron
energy  $\sqrt{s} = 1.8 \, {\rm TeV}$.  The predicted cross section
for  $p_\perp^{\, \rm jet} > 30 \, {\rm GeV/c}$ with $W \to e \nu$ or
$\mu \nu$ is $\approx 180 \, {\rm pb}$, which is large enough for an 
ample number of events to be collected in the CDF and D0  experiments
at the Fermilab Tevatron.  Similar physics can also  be studied in
$Z^0$ or direct photon + jet production.

At lowest order in QCD, $W + {\rm jet}$ production comes from three 
incoherent partonic subprocesses:
\begin{eqnarray}
\label{eq:type1} g q \to W q  \\
\label{eq:type2} q g \to W q  \\
\label{eq:type3} q \bar q \to W g 
\end{eqnarray}
The first two of these are ``Compton'' subprocesses that   are
distinguished from each other by the direction of their initial 
gluon, which is in the  $+ z$ direction for (\ref{eq:type1}) and $-
z$ for (\ref{eq:type2}).   The third ``annihilation'' subprocess has
a slightly larger cross section  than the sum of the Compton
subprocesses, because of the dominance of  quarks over gluons in the
parton distributions at the moderate values  of $x$ involved here.

``Partonometry'' in this instance is the ability to recognize which 
of the three hard subprocesses (\ref{eq:type1})--(\ref{eq:type3})
was  responsible for a given $W + {\rm jet}$ event. In section 2 we
will see that the subprocesses have very distinct radiation patterns.
These patterns are quite striking, but unfortunately the ability to
distinguish them is hampered by large fluctuations of the radiation
around these average distributions.  To model the fluctuations, as
well as the underlying background event and hadronization effects, 
we use the Monte Carlo program {\footnotesize HERWIG} \cite{herwig},
which has the correct  correlations included to produce the radiation
patterns that we seek. In section 3 we define some event variables
that are useful for  distinguishing the underlying subprocesses.   In
section 4 we describe the Monte Carlo simulation that we use and show
that it reproduces the radiation patterns that we expect, even after
the underlying event and hadronization have been included. Our main
results are in section 5, where we identify an observable of the soft
radiation which can be used to discriminate between the parton-level
subprocesses, and we show how this observable can be tested by
experiment.  Finally, in section 6 we give our conclusions.

\section{Radiation Patterns}

\begin{figure}
    \begin{center}
        \begin{tabular}{cc}
            \includegraphics[scale=0.7]{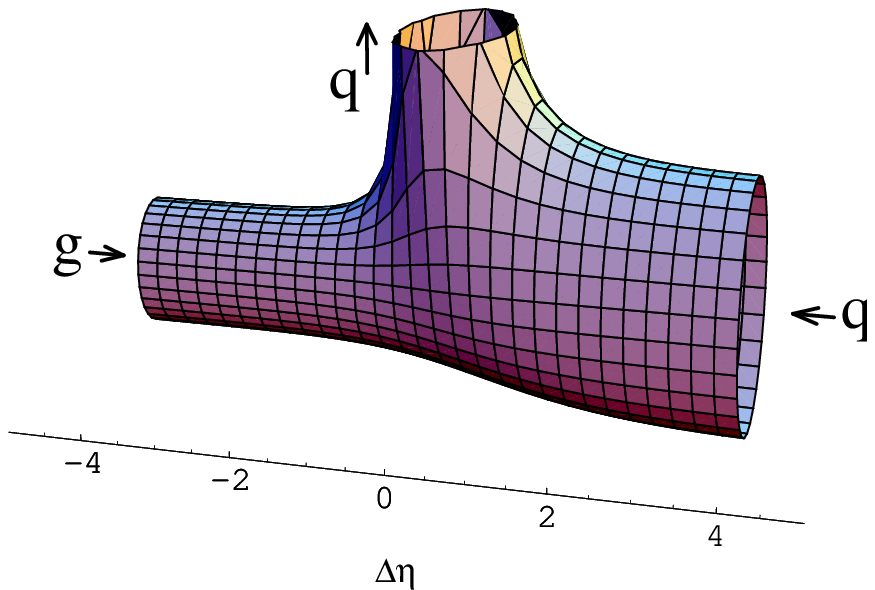} & 
            \includegraphics[scale=0.7]{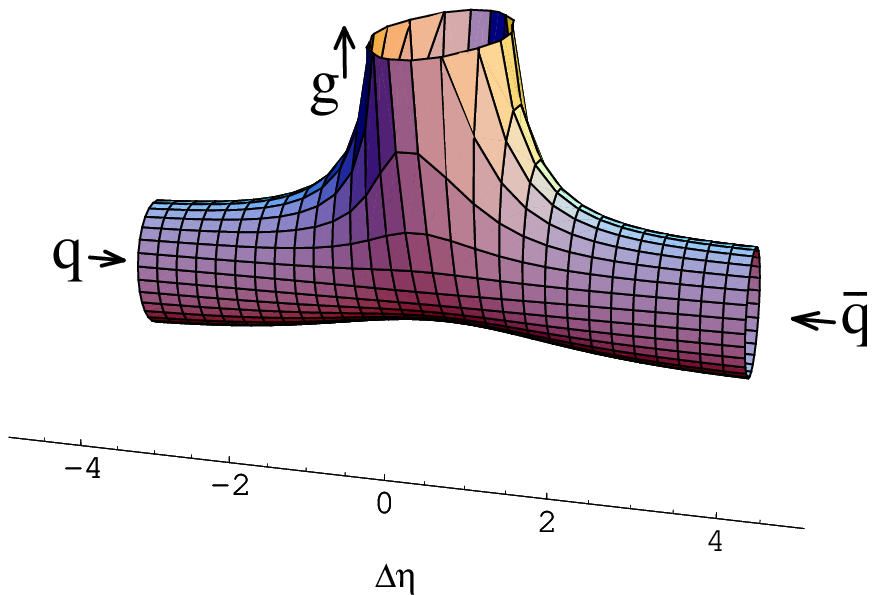} \\
            (a) & (b) \\
    	\end{tabular}
    \end{center}
    \caption{Radiation densities (a) $\rho_1$ and (c) $\rho_3$.
    $\rho_2$ can be obtained from $\rho_1$ by the replacement $\Delta
    \eta \to -\Delta \eta$.}
    \label{fig:pipe}
\end{figure}

In the large $N_c$ limit, soft gluon radiation from the three basic
subprocesses is proportional to \cite{khoze,dokshitzer}
\begin{eqnarray}
\label{eq:patt1} \rho_1 &=& 1 + \frac{\frac{1}{2} \, e^{+ \Delta \eta}} 
{\cosh{\Delta \eta} - \cos{\Delta \phi}}  \\
\, \null \nonumber \\  
\label{eq:patt2} \rho_2 &=& 1 + \frac{\frac{1}{2} \, e^{- \Delta \eta}} 
{\cosh{\Delta \eta} - \cos{\Delta \phi}} \\
\, \null \nonumber \\  
\label{eq:patt3} \rho_3 &=&  \frac{\cosh{\Delta \eta}} 
{\cosh{\Delta \eta} - \cos{\Delta \phi}}
\end{eqnarray}
for processes (1), (2), and (3), respectively.   The ``Lego'' 
variables here are defined relative to the jet location:  
\begin{equation}
    \Delta \eta \equiv \eta - \eta_{\rm jet}
\end{equation}
and
\begin{equation}
    \Delta \phi \equiv \phi - \phi_{\rm jet},
\end{equation}
with  $\eta = - \ln \tan \theta/2$ the pseudorapidity and $\phi$ the
azimuthal angle. The soft gluon radiation patterns are displayed in
cylindrical-coordinate plots in Fig.~\ref{fig:pipe}.  Here $\Delta \eta$
is along the cylindrical axis, $\Delta \phi$ is the azimuthal angle, and
the intensity of the radiation is proportional to the distance from the
cylindrical axis.  Note that these radiation patterns only depend on the 
coordinates relative to the jet and are completely independent of the 
direction of the $W$ boson.

There are three important gross features of the radiation patterns. 
A first obvious feature is that the radiation patterns  are singular
in the direction collinear to the outgoing jet. This is exhibited by
the fact that the surfaces in  Fig.~\ref{fig:pipe} blow up to
infinity as $\Delta \eta \to 0$, $\Delta \phi \to 0$.  In addition, 
the radiation from a gluon jet is stronger than radiation from  $q$
or $\bar q$ as expected:   $\rho_1 = \rho_2 = \frac{1}{2} \rho_3 \to 
1/[(\Delta \eta)^2 + (\Delta \phi)^2]$ for small $\Delta \eta\mbox{,
}\Delta\phi$.  

The second general feature is that radiation from the Compton
subprocesses is enhanced in the  direction of their initial gluon:
\begin{eqnarray}
\left.
\begin{array}{l}
\rho_1 \to 1 \\
\rho_2 \to 2 \\
\rho_3 \to 1
\end{array} \right\} \; \mbox{for} \; \Delta \eta \gg 1 \; ,
\quad
\left.
\begin{array}{l}
\rho_1 \to 2 \\
\rho_2 \to 1 \\
\rho_3 \to 1
\end{array} \right\} \; \mbox{for} \; \Delta \eta \ll 1 \; .
\label{eq:limits1}
\end{eqnarray}
This enhancement extends arbitrarily far away from the jet in the 
$(\eta,\phi)$ plane.  The large area over which the radiation pattern
can be integrated offers the prospect of observing it in a majority
of events, even though the observed radiation in any small region of
$(\eta,\phi)$ is subject to large fluctuations and is partially
masked by the ``background event''. Note that we use the variables
$\Delta \eta$ and $\Delta \phi$ rather than $\Delta R$ and $\beta$ of
Ref.~\cite{khoze}  because we have found regions distant from the jet
to be important. The relations between these variables are $\Delta
\eta = \Delta R \, \cos \beta$ and $\Delta \phi = \Delta R \, \sin
\beta$. 

The third important gross feature of the radiation pattern is the
suppression  of radiation from  the annihilation subprocess
(\ref{eq:type3}) at small $\Delta \eta$  on the side opposite to the
jet in azimuthal angle.   The extreme of this is 
\begin{eqnarray}
\left.
\begin{array}{l}
\rho_1 \to 1.25 \\
\rho_2 \to 1.25 \\
\rho_3 \to 0.5
\end{array} \right\} \; 
\mbox{for} \; \Delta \eta = 0, \Delta \phi = 180^\circ \; .
\label{eq:limits2}
\end{eqnarray}

\begin{figure}
    \begin{center}
    \begin{tabular}{cc}
      \includegraphics[scale=0.4]{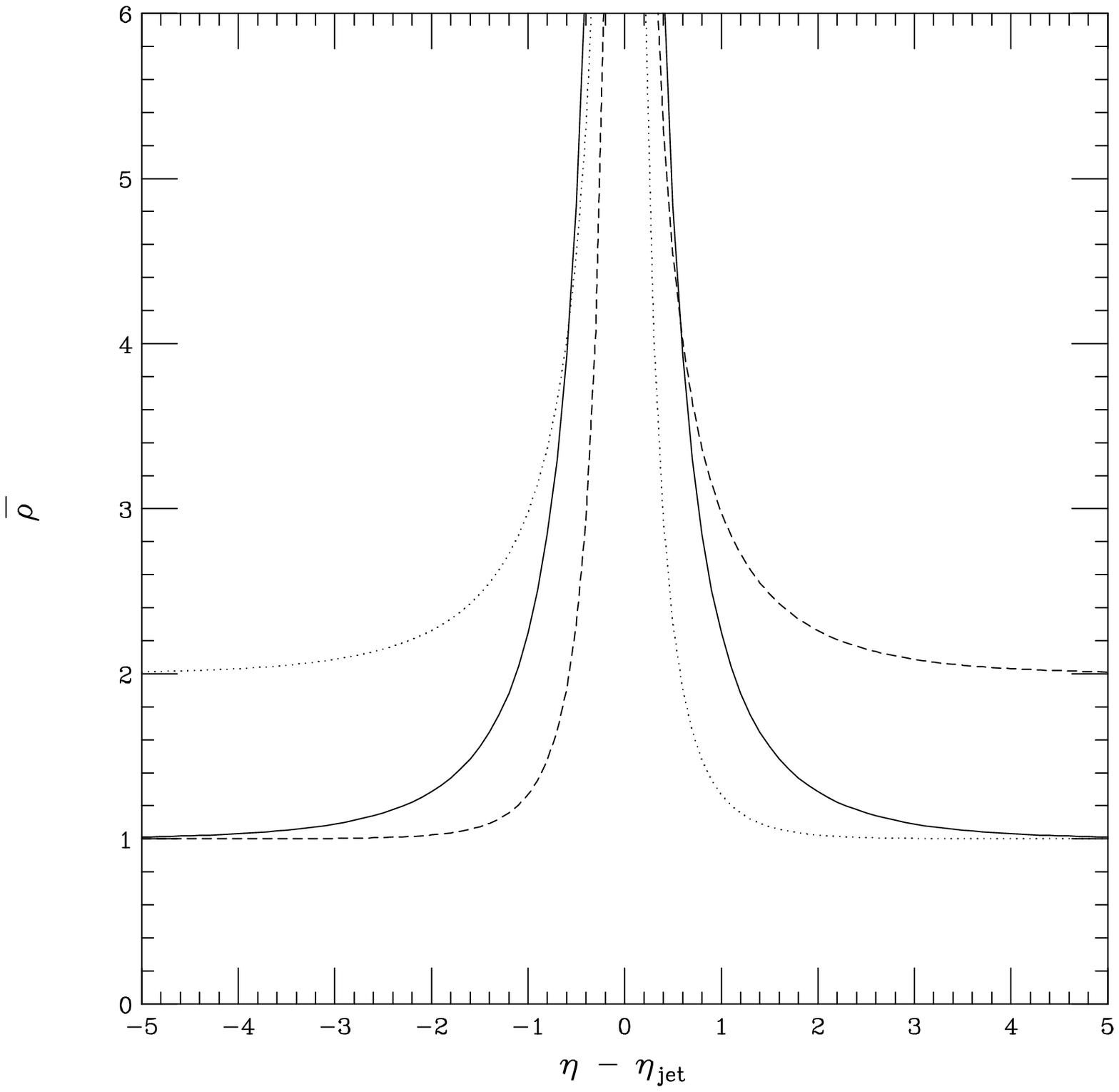} &
      \includegraphics[scale=0.4]{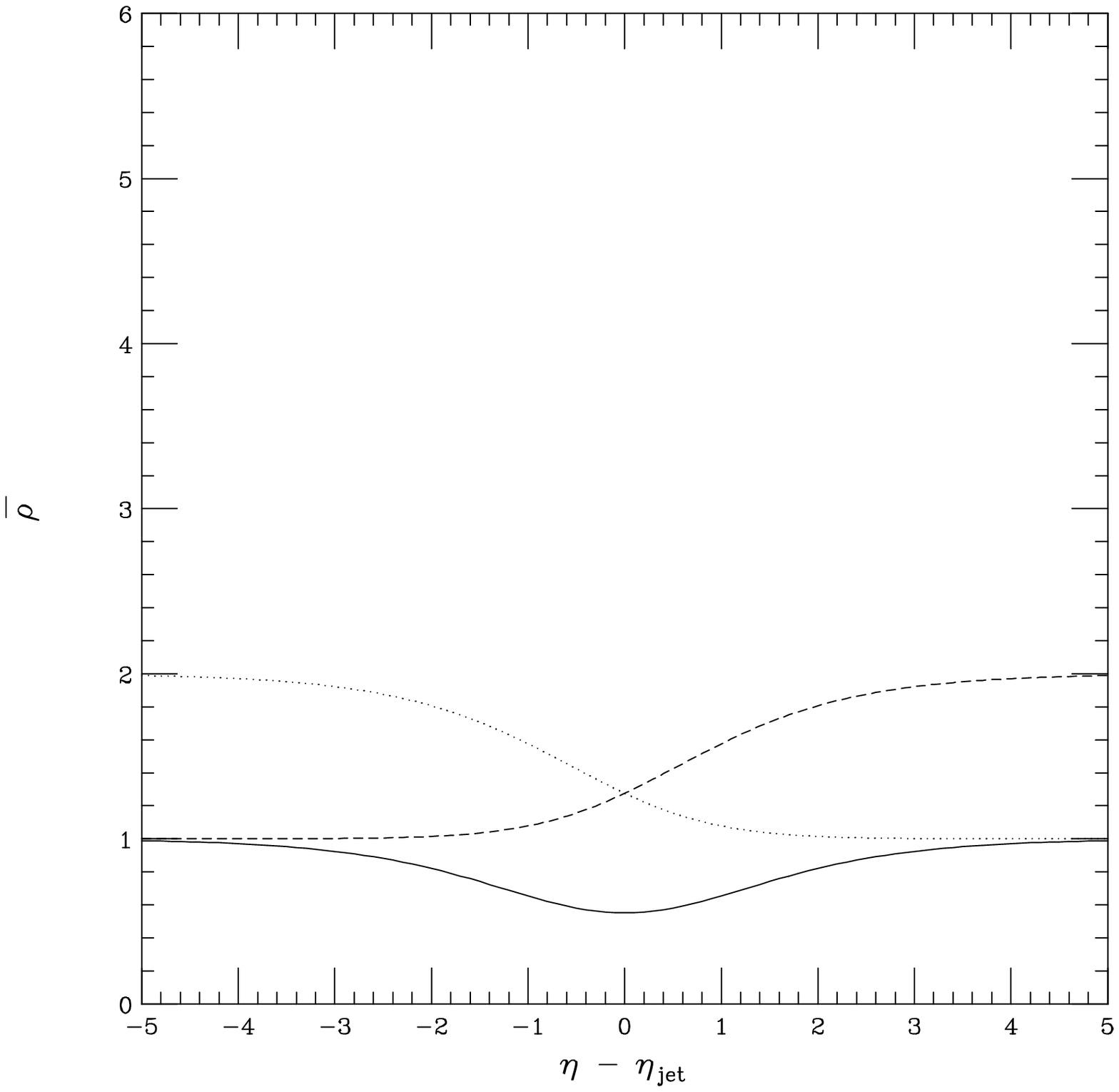} \\
      (a) & (b)
    \end{tabular}
    \end{center}
 \caption{
(a) Soft gluon radiation function averaged over strip 
$0 < |\Delta \phi| < 60^\protect\circ$.  
Dashed curve = subprocess (\protect\ref{eq:type1}), 
Dotted curve = subprocess (\protect\ref{eq:type2}), 
Solid curve = subprocess (\protect\ref{eq:type3}).
(b) Same, but for 
$120^\protect\circ < |\protect\Delta \protect\phi| < 180^\protect\circ$ 
}
\label{figure1}
\end{figure}

These features of the radiation patterns can be seen more quantitatively in 
Fig.~\ref{figure1}, 
which shows the $\Delta \eta$ distributions averaged over strips 
$0 < |\Delta \phi| < 60^\circ$ and 
$120^\circ < |\Delta \phi| < 180^\circ$.
Although we use the large $N_c$ limit throughout, we note that the
$1/N_c^2$ corrections are small and actually enhance the effects that we
are seeking.

\section{Event Variables}

\begin{figure}
    \begin{center}
    	\includegraphics[scale = 0.5]{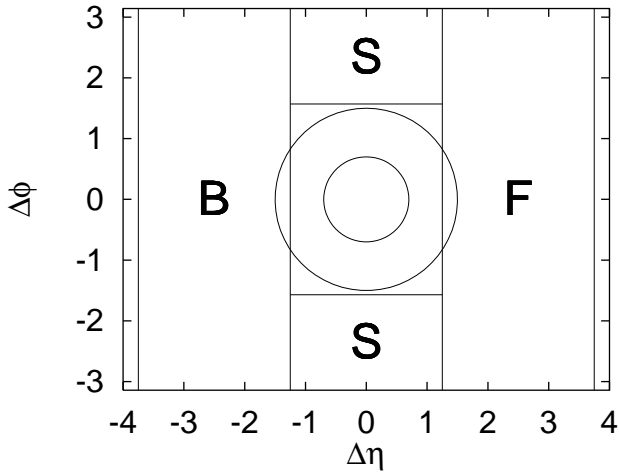}
    \end{center}
    \caption{Definition of regions F, B and S.}
    \label{fig:FBS}
\end{figure}

Before we look for soft radiation patterns in the data, we define
some convenient observable quantities in order to highlight the features
discussed above. We divide the region outside the jet into forward
(F), backward (B) and side (S) regions as in Fig.~\ref{fig:FBS}.
We then define the following scalar sums
\begin{equation}
F =\sum_{+1.25  \, <  \, \Delta\eta < \,  +3.75} |p_\perp| \; ,
\label{eq:sumf} 
\end{equation}
\begin{equation}
B = \sum_{-3.75  \, <  \, \Delta\eta \, < \,  -1.25} |p_\perp| \; ,
\label{eq:sumb} 
\end{equation}
and
\begin{equation}
S =\sum_{|\Delta\eta| < 1.25, \; |\Delta\phi| > \pi/2} |p_\perp| \; . 
\label{eq:sums}
\end{equation}

With these definitions, we would expect on average that subprocess
(1) ($ gq\rightarrow Wq$) would have an enhancement in F, subprocess
(2) ($ qg\rightarrow Wq$) would have an enhancement in B, and
subprocess (3) ($ q\bar q\rightarrow Wg$) would have a deficit in S.
We have investigated a number of observable quantities that use this 
information to distinguish between the underlying subprocesses.  The 
details are given in the appendix.  In the  main part of the text we
will focus on the quantity $|F-B|+S$. It is designed to be sensitive
to the differences between events with a final-state quark (1 and 2)
and events with a final-state gluon (3).  This is because quark-jet
events tend to have large forward-backward differences and relatively
large radiation in the side region, as we saw in the previous
section. Gluon-jet events, on the other hand, tend to have equal
forward and backward radiation, in addition to suppression in the
side region.

To test this observable, it will be helpful to have another quantity 
sensitive to the differences between the various subprocesses that
{\em does not} depend on the properties of the soft radiation.
Various ways to measure the radiation inside a jet cone have  been
invented in order to discriminate between quark jets and gluon 
jets.  They have been validated both by using Monte Carlo simulation 
\cite{qgsep,xvar}  and experiment \cite{opal}.   One suitable variable
\cite{xvar} can be defined by dividing the jet  cone of radius $R =
0.7$ in $(\eta,\phi)$ into $0.1 \times 0.1$ cells,  as is typical of
calorimetric detectors.  The jet $p_\perp$ can  be defined as the
total $p_\perp$ detected in all cells in the cone. By sorting these
cells in order of their $|p_\perp|$, we define the  variable 
\begin{equation}
X = \mbox{the 
minimum number of cells needed to include all but 
$\sqrt{|p_\perp|^{\rm tot}}$ of the total $|p_\perp|^{\rm tot}$},
\end{equation}
where $p_\perp$ is given in units of GeV/c. This definition makes the
$X$ distribution roughly independent of  jet $p_\perp$. $X$ can be
defined to include a fractional part given by linear or  logarithmic
interpolation.

\section {Simulation}
\label{sec:simulation}

The distinctive radiation patterns of  Fig.~\ref{figure1} are
predicted by  the soft gluon approximation.  However, on an
event-by-event basis there will be large fluctuations around these
average distributions.   We must also ask to what extent  these
patterns survive finite-energy effects and the  non-perturbative
effects of hadronization, whereby colored partons combine to form
physical hadrons.  Many of those hadrons are  resonances that undergo
strong decay before reaching the detector,  further broadening
structures in the $(\eta,\phi)$ plane.  The  notion of local
parton-hadron duality (LPHD) \cite{lphd}, which  is supported by
experiment \cite{ochs}, suggests however that the  patterns should
survive reasonably well.

For quantitative predictions, we simulate  $\bar p p \to W^\pm + {\rm
jet}$ using the  {\footnotesize HERWIG 5.8} \cite{herwig}  Monte
Carlo event generator, which contains the color coherence  effects we
seek and embodies LPHD in its cluster  formation and decay model.   
We make a rather tight cut on the jet transverse momentum  ($30 <
p_\perp^{\, \rm jet} < 40$ GeV/c)  in order to  simplify the study.  We
will also generally make a  cut $|\eta_{\rm jet}| < 0.5$ on the jet
rapidity.  These  cuts are not essential, and it  will be desirable
to loosen them for actual data analysis to  improve the statistics. 
In the experiment, the  cut on minimum $p_\perp^{\, \rm jet}$ should
be replaced by  a cut on $p_\perp^{\, \rm W}$ to avoid biases that
depend on  jet structure.  The effects we  seek are somewhat cleaner
at higher $p_\perp^{\, \rm jet}$,  but the cross section falls rather
rapidly at large  $p_\perp^{\, \rm jet}$. 

\begin{figure}
    \begin{center}
    \begin{tabular}{cc}
      \includegraphics[scale=0.4]{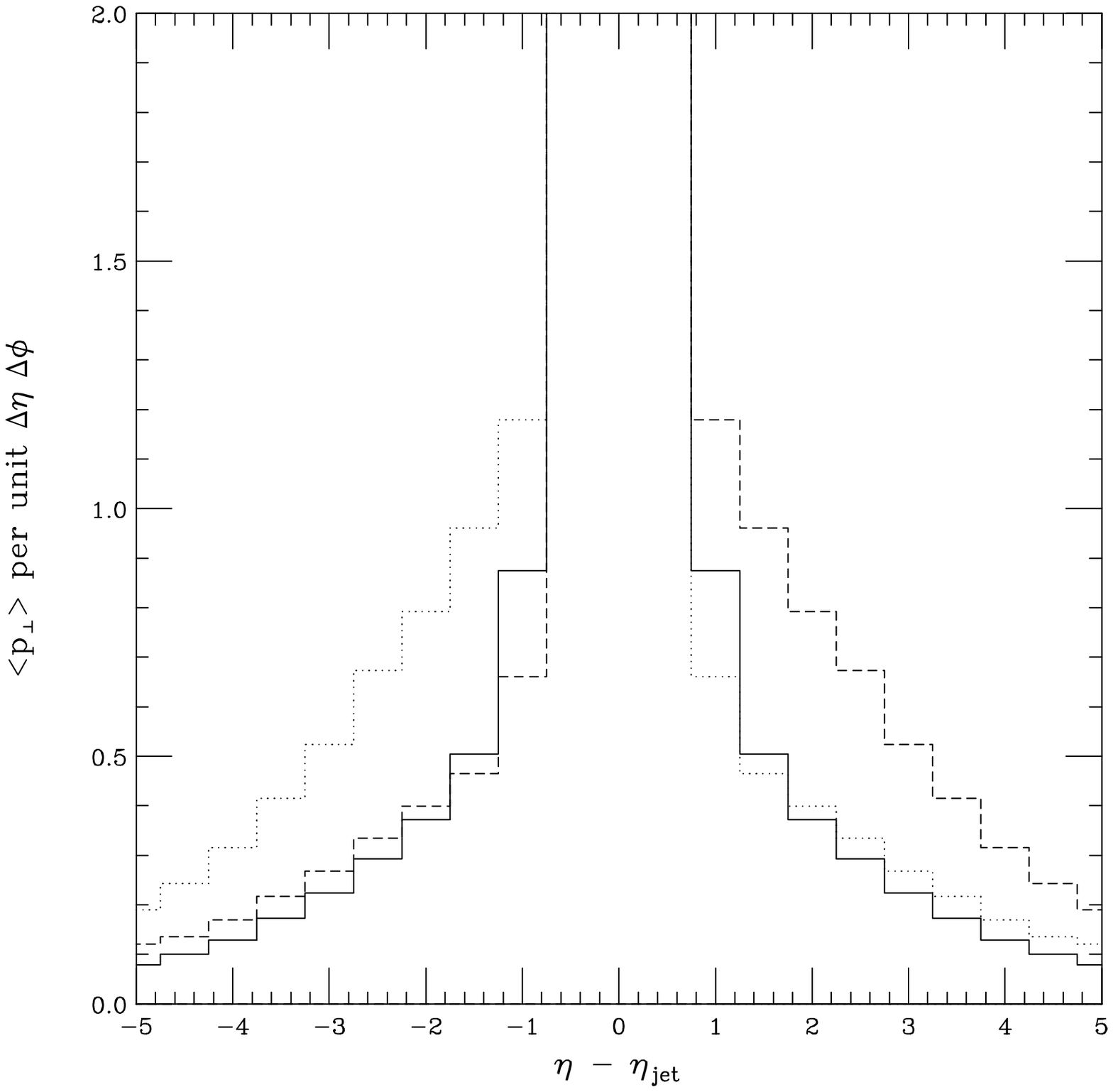} &
      \includegraphics[scale=0.4]{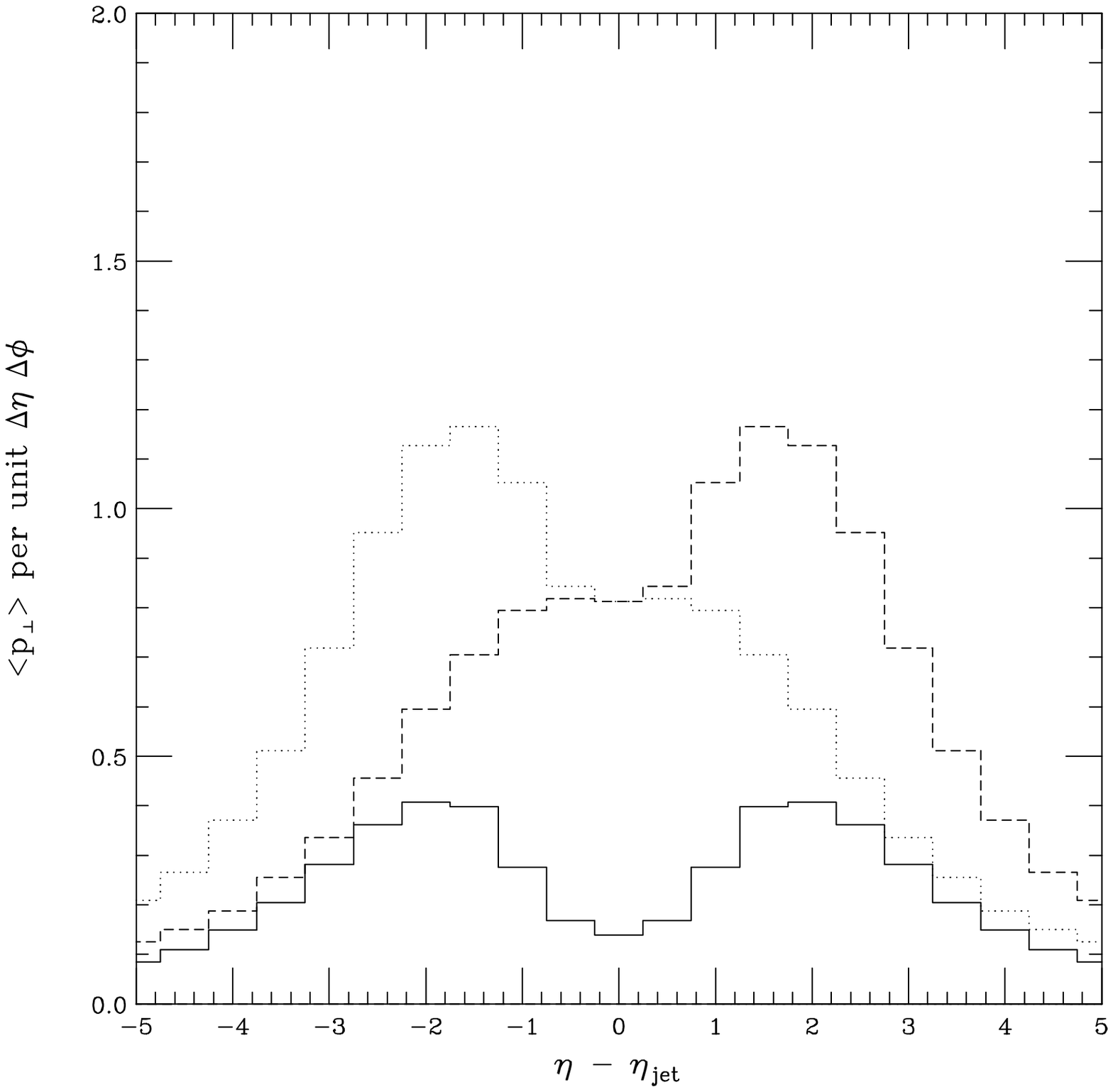} \\
      (a) & (b)
    \end{tabular}
    \end{center}
\caption{
(a) Average $p_\perp$ per unit area in $(\eta, \phi)$  
for simulated events with $|\eta_{\rm jet}| < 0.5$,
in the 
region $0 < |\Delta \phi| < 60^\circ$ 
(cf. Fig.~\protect\ref{figure1}), 
``background event'' off. 
(b) Same, but for 
$120^\circ < |\Delta \phi| < 180^\circ$.
}
\label{figure4}
\end{figure}

\begin{figure}
    \begin{center}
    \begin{tabular}{cc}
      	\includegraphics[scale=0.4]{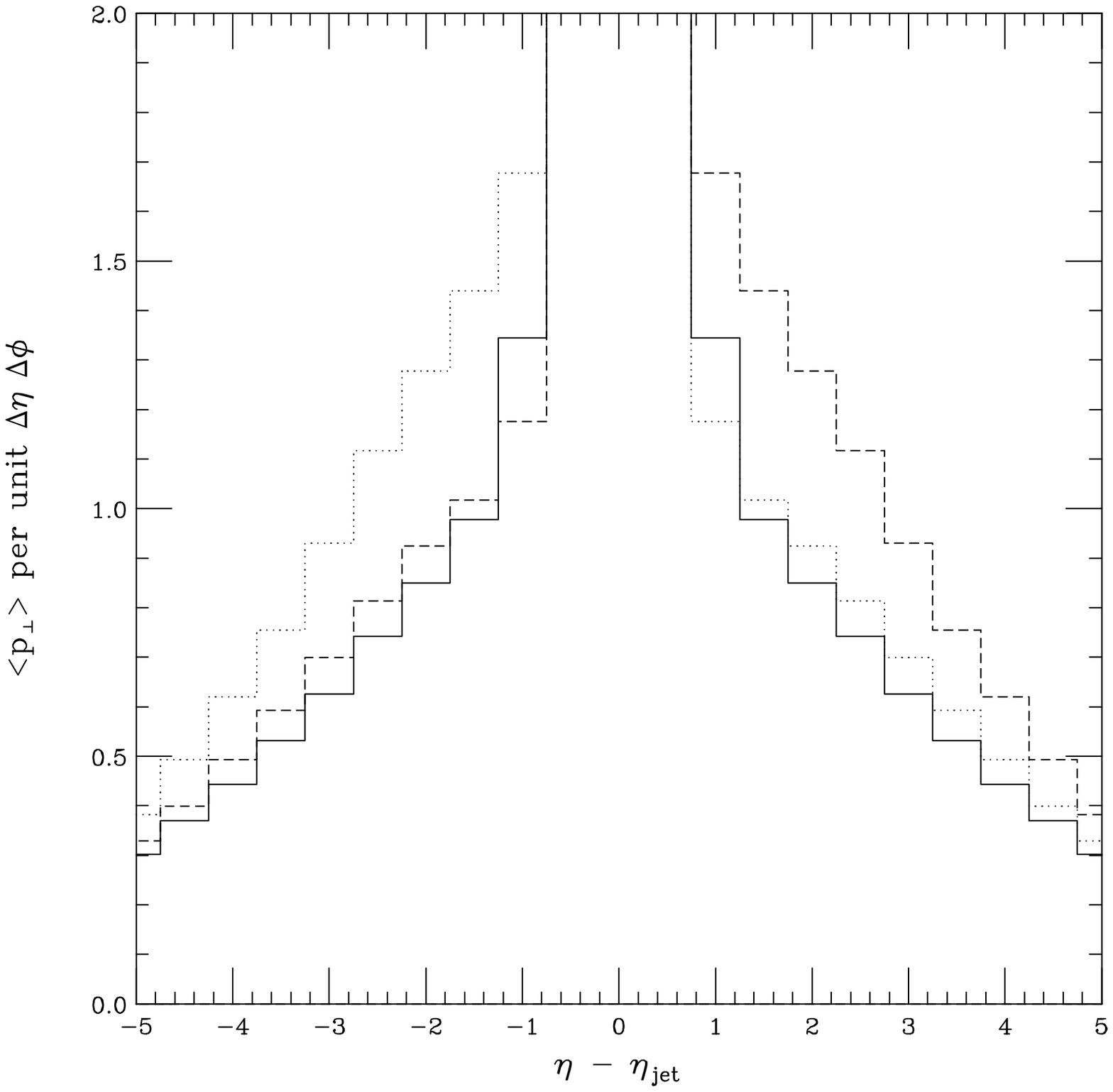} &
      	\includegraphics[scale=0.4]{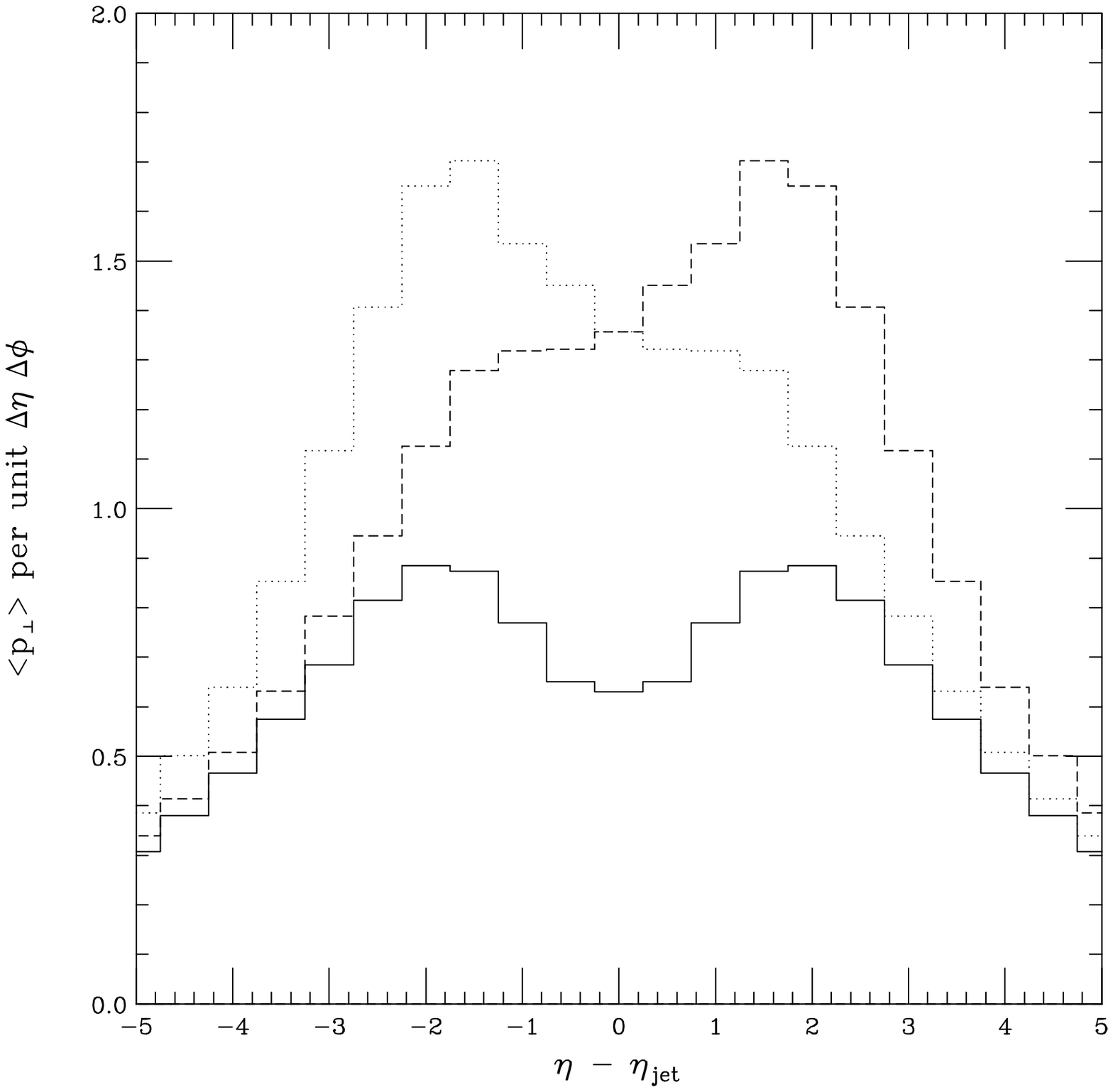} \\
      (a) & (b)
    \end{tabular}
    \end{center}
\caption{
(a) Average $p_\perp$ per unit area in $(\eta, \phi)$  
for simulated events with $|\eta_{\rm jet}| < 0.5$,
in the 
region $0 < |\Delta \phi| < 60^\circ$ 
(cf.\ Fig.~\protect\ref{figure1}), 
``background event'' on. 
(b) Same, but for 
$120^\circ < |\Delta \phi| < 180^\circ$.
}
\label{figure7}
\end{figure}

Results from the simulation are shown in Fig.~\ref{figure4}  for the 
same two strips in $\Delta \phi$ as in  Fig.~\ref{figure1}.   The
quantity plotted is  $\langle p_\perp \rangle$ in GeV/c per unit
$\Delta \eta \; \Delta \phi$ at  $|\eta_{\rm jet}| < 0.5\,$.  A
suppression at large   $\eta$ is apparent, so that the ``plateau'' is
not flat.  But otherwise, the soft gluon pattern  survives very
well.  In particular the roughly $2:1$ enhancement  of subprocesses
(\ref{eq:type1}) and (\ref{eq:type2})  in the directions of their
incoming  gluon at large $\Delta \eta$, and the  suppression of
subprocess (\ref{eq:type3}) at  $\{\eta \approx \eta_{\rm jet}, 
\Delta \phi \approx 180^\circ\}$ remain.  In generating 
Fig.~\ref{figure4}, the {\footnotesize HERWIG} option of a
contribution from  a minimum-bias--like ``background event'' was
turned off.  The result  of turning it on is shown in
Fig.~\ref{figure7}.  The background event adds  extra radiation to
each subprocess, so that the differences are largely  preserved,
while the ratios are substantially diluted.

\section {Observing the radiation patterns}
\label{sec:observing}

Each of the partonic subprocesses (\ref{eq:type1})--(\ref{eq:type3})
is  predicted to have its  distinctive radiation pattern, as we have
seen.   Unfortunately, these radiation patterns cannot be observed 
directly, because for real events --- unlike the simulated ones ---  
one does not know which of the three subprocesses is responsible  for
any given event.  To establish this physics with a testable 
prediction, we must study a global correlation that is  based on the
fact that any particular event is caused by a single  one of the
three production subprocesses.  Therefore, a typical event should
have {\it either} enhanced radiation at  $\eta \gg \eta_{\rm jet}$;
{\it or}  enhanced radiation at  $\eta \ll \eta_{\rm jet}$; {\it or}
reduced  radiation at $\eta \approx \eta_{\rm jet}$,  $\Delta \phi
\approx 180^\circ$. Looking for these  characteristics of the
radiation in an event-by-event basis  is also what is needed to
achieve the promise of  ``partonometry'' \cite{khoze}. Of course, in
practice the underlying subprocess  can only be determined with some
probability, due to fluctuations in the soft radiation patterns.

There are three necessary steps to partonometry. First, we must
identify an observable function of the soft radiation that
discriminates between partonic subprocesses. Second, we need to be able
to test the prediction experimentally. Finally, we need a
quantitative measure of the discrimination efficiency of our
observable. The second step requires comparison with another
event observable that {\em does not} depend on the properties of the
soft radiation in the event. An observable based on information inside
the jet cone is the natural choice.

In what follows we give a concrete example of partonometry using the
soft radiation observable $|F-B|+S$. We also use the jet observable
$X$ as a cross check of this discriminator. These may or may not
prove to be the best variables when detector effects are included.
Nonetheless, the procedure should be applicable to a wide variety of
observable quantities.

\begin{figure}
    \begin{center}
      \includegraphics[scale=0.4]{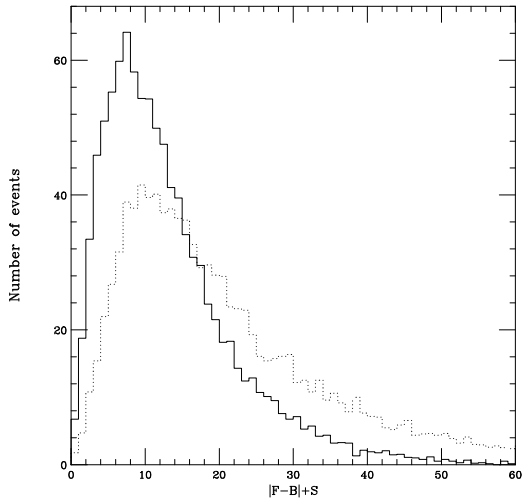}
    \end{center}
\caption{
Probability distribution for $|F-B| + S$.
Solid curve: quark jets (from subprocess (\protect\ref{eq:type3}));
dotted curve: gluon jets (from subprocesses (\protect\ref{eq:type1}) and 
(\protect\ref{eq:type2})).
}
\label{fig:FmBpS}
\end{figure}

\begin{figure}
    \begin{center}
      \includegraphics[scale=0.4]{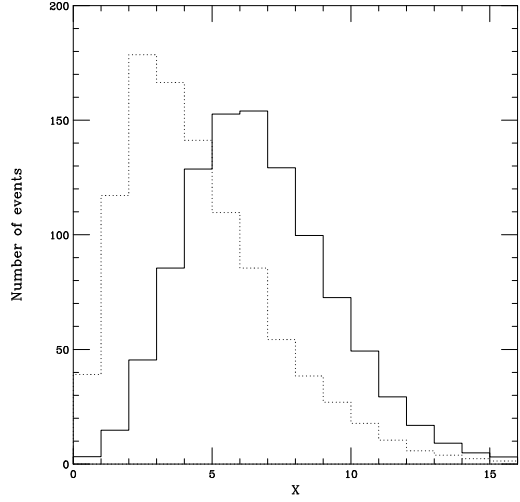}
    \end{center}
\caption{
Probability distribution for $X$ defined as the minimum number  of
cells in the jet cone $R < 0.7$ required to contain all but 
$\protect\sqrt{|p_\perp|^{\rm tot}}$ of the total jet $|p_\perp|^{\rm
tot}$ in the cone, where $p_\perp^{\rm
tot}$ is given in GeV/c.  Solid curve: quark jets (from subprocess
(\protect\ref{eq:type3})); dotted curve: gluon jets (from subprocesses
(\protect\ref{eq:type1}) and  (\protect\ref{eq:type2})).
}
\label{fig:X}
\end{figure}

The predictions from {\footnotesize HERWIG} for $|F-B|+S$ are shown in
Fig.~\ref{fig:FmBpS}.
The differences between the curves satisfy our first criterion above.
The predictions from {\footnotesize HERWIG} for the jet variable $X$ are shown 
in
Fig.~\ref{fig:X}.
There is a clear difference between the quark  jets (subprocess
(\ref{eq:type3})) and the gluon jets  (subprocesses (\ref{eq:type1})
and (\ref{eq:type2})), with the latter  tending toward larger values
of $X$, i.e., gluon jets are fatter. The correlations between
Figs.~\ref{fig:FmBpS} and \ref{fig:X} provide the qualitative and
quantitative tests for our soft radiation predictions.

\begin{figure}
    \begin{center}
    \includegraphics[scale=0.4]{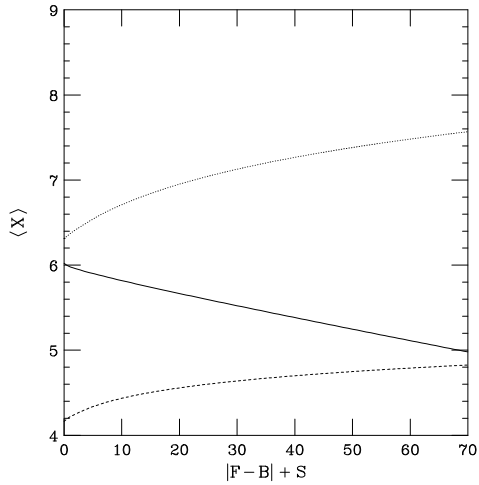}
    \end{center}
    \caption{Mean value of $X$ as a function of $|F-B|+S$.}
    \label{fig:xfbs}
\end{figure}

\begin{figure}
    \begin{center}
    \includegraphics[scale=0.4]{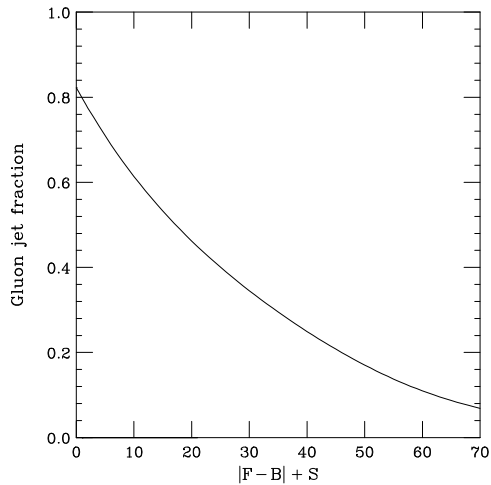}
    \end{center}
    \caption{Fraction of jets which are gluon jets, $g$, as a function
    of $|F-B|+S$.}
    \label{fig:gluonfrac}
\end{figure}

To observe the correlation, we divide the data sample into bins of
$|F-B|+S$ and then calculate the mean value of $X$ in each bin.  The 
result is shown in
Fig.~\ref{fig:xfbs}.
For a pure sample of gluon jets or quark jets, the value of $\langle
X \rangle$ increases mildly with $|F-B|+S$, indicating that $\langle
X \rangle$ increases with the amount of soft radiation, as one would
expect. However, we predict that the combined data will show a {\em
decrease} in $\langle X \rangle$ with increasing $|F-B|+S$. This is
an unambiguous qualitative test of the  soft radiation patterns. 
The decrease comes about because as 
$|F-B|+S$ increases, the likelihood that the outgoing jet is a quark
increases, reducing the value of $\langle X \rangle$.  This is shown 
in 
Fig.~\ref{fig:gluonfrac}, where we plot the fraction of outgoing gluon 
jets as a function of $|F-B|+S$, as calculated using {\footnotesize HERWIG}.  
The gluon
or quark jet fraction can be made as high as 80\% by focusing on 
events with very large or small values of $|F-B|+S$, albeit at the 
expense of statistics.
Even stronger discrimination between the processes can be achieved by 
using both variables $|F-B|+S$ and $X$ in tandem.

\section {Conclusions}
\label{sec:conclusions}

Soft radiation patterns in $W+ \mbox{jet}$ production provide a novel
insight into the structure of QCD.  In this paper we have identified 
the most important structures in the radiation patterns, and we have 
defined an observable, $|F-B|+S$, that can be used to distinguish
the  three partonic subprocesses (\ref{eq:type1})--(\ref{eq:type3}).
We have also shown how this observable can be cross-checked against 
a second observable, $\langle X \rangle$, which relies only on
information  from inside the jet cone.   The prediction, shown by the
solid curve in Fig.~\ref{fig:xfbs}, is a  striking one:  we predict a
negative correlation between the amount  of radiation inside the jet
cone and the amount of radiation outside  it.  This arises solely
because more radiation outside the cone is  associated with a higher
probability of the jet being a quark jet, as  shown in
Fig.~\ref{fig:gluonfrac}.  The negative correlation is otherwise 
opposite to the naive expectation that the radiation  outside the
cone should rise with the radiation inside -- as is confirmed for 
quark jets alone or gluon jets alone in  Fig.~\ref{fig:xfbs}. The
ability to discriminate between the underlying soft radiation 
patterns is potentially useful as a tool for identifying and studying
parton-level subprocesses.  One could imagine using this information
both as a means to separate the  gluon from the quark parton
distribution functions in the initial  state and as a method to study
the differences between quark and gluon jets in the final state. 

We emphasize again that the procedure we have outlined for  utilizing
and testing the soft-radiation effects could be followed for  a
variety of soft-radiation and jet observables.  Further work may be  
needed in particular to optimize the suggested observables to suit 
detector limitations.  In addition, the details presented here rely
on the use of {\footnotesize HERWIG}  for  producing the
hadronization effects and the background event.   Information on jet
properties taken directly from experiment  \cite{opal} may be used to
lessen this dependence on the Monte Carlo  simulation.  The important
point is that the soft radiation effects  should remain and can be
discerned. Observing these patterns is both interesting in its own
right and as a novel  probe of the underlying partonic structure of
QCD.

\section*{Acknowledgments}

We thank Harry Weerts and Wu-Ki Tung for helpful discussions. This
work was supported in part by U.S. National Science Foundation grant
number PHY--9507683.

\section*{Appendix}

In the text we have used the quantity $|F-B|+S$ to discriminate
between the soft radiation patterns from the different event types.
Many other possibilities are available. We give some of the
alternatives here. 

One logical quantity to investigate is the asymmetry
\begin{equation}
    A \equiv \frac{F-B}{F+B}.
\end{equation}
In the limit where the forward and backward regions extend to
infinity, Eqs.~(\ref{eq:patt1}--\ref{eq:patt3}) predict the averages
$\langle A \rangle = +1/3,~-1/3\mbox{ and }0$ for subprocesses 1, 2,
and 3, respectively. This prediction does not include such real-world
effects as hadronization and the underlying event, which tends to
increase both $F$ and $B$.  

\begin{figure}
    \begin{center}
      \includegraphics[scale=0.4]{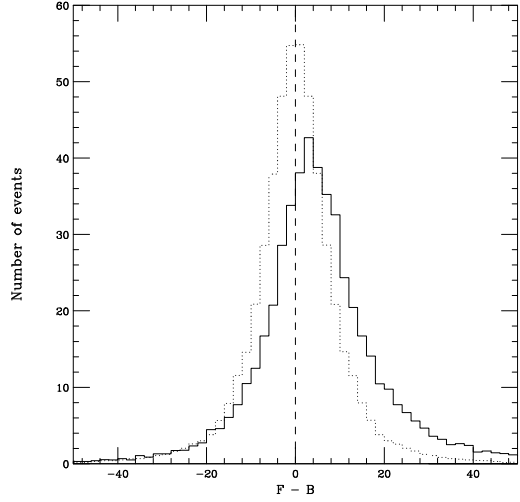}
    \end{center}
\caption{
Normalized distributions of $F-B$ for events with $|\eta_{\rm jet}| < +0.5$.
  Solid curve:  quark jets (from subprocess (\protect\ref{eq:type1})), 
dotted curve:  gluon jets (from subprocess (\protect\ref{eq:type3})).  
The distribution for quark jets from
subprocess (\protect\ref{eq:type2}) would be the same as for subprocess
(\protect\ref{eq:type1}) 
with $F - B$ replaced by $-(F-B)$.
}
\label{figure11}
\end{figure}

The forward-backward difference  
$F-B$ itself is less sensitive to the underlying event.
The histogram in Fig.~\ref{figure11} shows the probability
distribution for $F - B$. 
The peak is at  $F - B > 0$ for subprocess (\ref{eq:type1}), as expected
from the  average antenna patterns.  The fraction of events of this type 
with $F - B > 0$ is $67\%$.  In other words, if one  attempts to
discriminate between equal numbers of  subprocess (\ref{eq:type1}) and 
subprocess (\ref{eq:type2}) on the basis of $F - B$, the correct 
assignment is made $67\%$ of the time.  

\begin{figure}
    \begin{center}
      \includegraphics[scale=0.4]{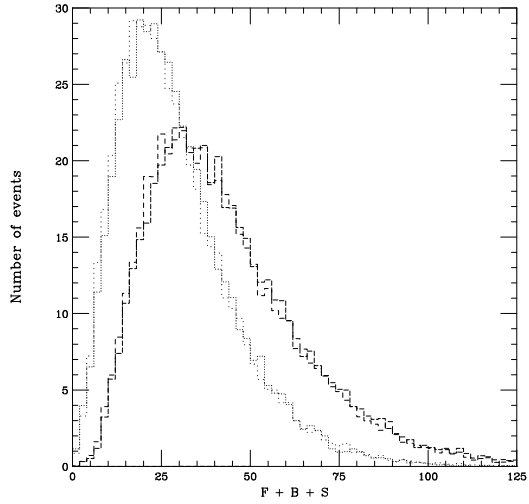}
    \end{center}
\caption{
Normalized distributions of $F + B + S$.  
Dashed curve: quark jets (from subprocess (\protect\ref{eq:type1})
or 
(\protect\ref{eq:type2})), dotted curve:  gluon jets (from subprocess
 (\protect\ref{eq:type3})).  Normal 
dots and dashes are for $|\eta_{\rm jet}| < 0.5$; sparse ones are for 
$1 < |\eta_{\rm jet}| < 2$.
}
\label{figure10}
\end{figure}

A second quantity that can be considered is just the total amount of 
soft radiation $F+B+S$.
The histogram in Fig.~\ref{figure10} shows the probability distributions
for $F+B+S$ for two different ranges of the rapidity of the outgoing
jet. The distributions for $0.5 < |\eta_{\rm jet}| < 2$ are almost identical 
to those for more central jets $|\eta_{\rm jet}| < 0.5\,$, so in 
analyzing experiments, a large range of $\eta_{\rm jet}$ will be 
usable.

The most important property of a soft-radiation observable is the 
ability to distinguish events with outgoing gluon jets (subprocess 
(\ref{eq:type3})) from events with outgoing quark jets (subprocesses 
(\ref{eq:type1}) and (\ref{eq:type2})).  To quantify this ability for 
a given observable $\cal O$, we consider a sample of events that contains 
equal numbers of outgoing gluon and quark jets.  This is close to 
the actual predicted mix.  We then identify the 50\% of events with 
the largest values of the observable $\cal O$ as quark-jet events and the 
50\% of events with the smallest values of $\cal O$ as gluon-jet events.  
The percentage of correctly-assigned events is then a measure of the 
discrimination power of the observable $\cal O$ at 100\% efficiency.  

With this definition, we find a discrimination power for $|F-B|$ of
60\%.  As expected, the forward-backward difference is better at 
distinguishing subprocess (\ref{eq:type1}) from subprocess (\ref{eq:type2}) 
than it is at distinguishing either from subprocess (\ref{eq:type3}).
The total amount of radiation $F+B+S$ does better, with a 
discrimination power of 63.6\%. The quantity $|F-B|+S$ used in the body 
of the text improves 
upon this slightly further, having a discrimination power of 64.0\%.  
We emphasize that these numbers are given at 100\% efficiency.  By 
rejecting events with middling values for the observable, one can 
increase the discrimination power at a cost of decreased efficiency.

We have also explored other variations on the basic observables above,
including:
\begin{enumerate}
\item
Ignoring contributions from 
cells of size $0.1 \times 0.1$ in $(\eta,\phi)$ that receive less than 
$0.1$, $0.2$, or $0.3 \, {\rm GeV/c}$
of transverse momentum; 
\item
Using multiplicity or charged multiplicity in place of the scalar sum 
of $|p_\perp|$ as the measure of radiation intensity;
\item
Generalizing the observables to include $n\ge 3$ regions in
$(\Delta \eta, \Delta \phi)$, each which may be assigned a different weight.
\end{enumerate}
These variations were found to be useless.  In particular, 
variation (1) might have been expected to help by reducing the effect of the 
background event.   On the contrary, however, provided   {\footnotesize
HERWIG}'s model for the background event is correct,  this is more
than offset by a simultaneous reduction in the  accuracy of measuring
the signal.  Variation (3) was  explored extensively by using the
Monte Carlo data itself.  Cutting   the $(\Delta \eta, \Delta \phi)$
plane into  $n = 3 - 8$ discrete regions allowed each event to be
represented by  an $n$-tuple vector consisting of the transverse
momentum deposited  into each of the $n$ regions.  One could then
attempt to classify a given event, whose underlying subprocesses is
taken to be unknown, by its closest resemblance to other events
whose underlying subprocess is taken to be known, using a suitable
metric.  This procedure amounts to an idealization of what a neural
network approach would only be able  to approximate.  However, it was not
found to do any better than the classification based on $F$,
$B$, and $S$.

\begin{figure}
    \begin{center}
      \includegraphics[scale=0.4]{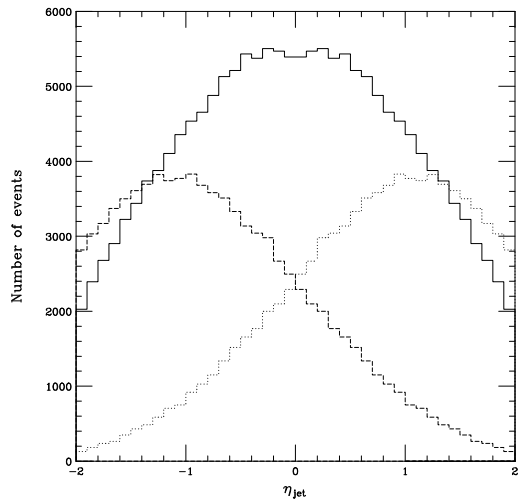}
    \end{center}
\caption{
Distributions of jet rapidity:  
dashed curve = subprocess (\protect\ref{eq:type1}), 
dotted curve = subprocess (\protect\ref{eq:type2}), 
solid curve = subprocess (\protect\ref{eq:type3}).
}
\label{figure12}
\end{figure}

For conceptual simplicity, we have concentrated on the region  $-0.5
< \eta_{\rm jet} < 0.5\,$.  To increase statistics, this region  can
be extended to $-2 < \eta_{\rm jet} < 2$ with very little change  in
the symmetric distributions such as $F + B + S$ or
$|F - B| + S$. Meanwhile, the ability to distinguish between subprocess
(\ref{eq:type1}) and subprocess  (\ref{eq:type2}) on the basis of the
antisymmetric quantity $F-B$  actually improves with increasing
$\vert \eta_{\rm jet}\vert$ because  of an accidental circumstance: 
the general radiation  pattern peaks around $\eta = 0$ for
kinematical reasons.  This  tendency increasingly favors negative
$F-B$ as $\eta_{\rm jet}$ becomes  positive; but subprocess
(\ref{eq:type2}) increasingly dominates over  subprocess
(\ref{eq:type1}) in that limit because of parton  distribution
effects, as is shown in Fig.~\ref{figure12}. 


\end{document}